\documentclass[pra,twocolumn,showpacs,floatfix]{revtex4}

\usepackage{amsmath,amssymb,amsthm}
\usepackage{times}
\usepackage{graphicx}
\usepackage{bm}

\newcommand{\leftindex}{\alpha}
\newcommand{\rightindex}{\beta}
\newcommand{\WFini}{\vert \psi_{ini}\rangle}
\newcommand{\WFtime}{\vert \psi(\tau)\rangle}

\newcommand{\EigenvO}{\vert\Psi_\leftindex^0\rangle}
\newcommand{\Eigenvleft}{\vert\Psi_\leftindex\rangle}

\begin{document}

\title{Quantum quenches and thermalization in one-dimensional fermionic systems}

\author{Marcos Rigol} 
\affiliation{Department of Physics, Georgetown University, Washington, DC 20057, USA} 


\begin{abstract}
We study the dynamics and thermalization of strongly correlated fermions in finite 
one-dimensional lattices after a quantum quench. Our calculations are performed
using exact diagonalization. We focus on one- and two-body observables such as the momentum 
distribution function [$n(k)$] and the density-density structure factor [$N(k)$], 
respectively, and study the effects of approaching an integrable point.
We show that while the relaxation dynamics and thermalization of $N(k)$ for fermions
is very similar to the one of hardcore bosons, the behavior of $n(k)$
is distinctively different. The latter observable exhibits a slower relaxation 
dynamics in fermionic systems. We identify the origin of this behavior, which
is related to the off-diagonal matrix elements of $n(k)$ in the basis of the
eigenstates of the Hamiltonian. More generally, we find that thermalization occurs far 
away from integrability and that it breaks down as one approaches the integrable point.
\end{abstract}

\maketitle


\section{Introduction}
\label{Sec:intro}

In recent years, the study of the nonequilibrium dynamics of isolated quantum systems
has attracted a great deal of attention. The main motivation for these studies lies behind the 
spectacular success that experimentalists have achieved in trapping and manipulating ultracold 
quantum gases. This has allowed them to, for example, load ultracold bosons in optical lattices 
and study the collapse and revival of the matter wave interference after quenching the interaction 
strength from deep in the superfluid regime into deep in the Mott insulating regime \cite{greiner02b},
and to observe the damping of the dipole oscillations when the center of mass of the gas was displaced away
from the center of the trap \cite{stoferle04,ott04a,fertig05}.

Since these atomic gases are trapped, cooled, and manipulated in a very high vacuum, i.e.,
in no contact with any kind of thermal or particle reservoir, a fundamental question that arises 
is whether after a sudden perturbation the system will relax to thermal equilibrium (an equilibrium 
in which observables are described by standard statistical 
mechanical ensembles). The usual assumption is that thermalization occurs in general. However, recent 
experiments in one-dimensional (1D) geometries (created by a deep two-dimensional optical lattice) 
have failed to observe relaxation to a thermal distribution after a quench \cite{kinoshita06}. 
The absence of thermalization can be understood in the very special regime in which the system 
is at an integrable point 
\cite{rigol07STATa}. Interestingly, in Ref.\ \cite{kinoshita06} the gas was perturbed away from  
integrability and thermalization did not occur. Other experiments in 1D geometries 
(generated by an atom chip) have reported the indirect observation of thermalization 
\cite{hofferberth07}. In the latter case, the transverse confinement was not as strong as the 
one induced by the optical lattice in Ref.\ \cite{kinoshita06}. The question of whether strict 
one dimensionality could affect the outcome of the relaxation dynamics has not been fully addressed
experimentally.

The particular case in which the system is at an integrable point lends itself to combined analytical
and numerical studies, which have allowed theorists to show that in that case the relaxation 
dynamics after a quench lead to nonthermal equilibrium distributions of few-body observables 
\cite{rigol07STATa}. In addition, a generalization of the Gibbs ensemble, which takes into 
account the conserved quantities that make the system integrable, has been shown to 
successfully describe experimental observables after relaxation \cite{rigol07STATa,rigol06STATb}. 
Further analytical and numerical works have analyzed the relevance of the generalized Gibbs 
ensemble to different models and observables and addressed its limits of applicability 
\cite{cazalilla06,calabrese07a,cramer08a,barthel08,eckstein08,kollar08,cramer08b,cramer08c,rossini09,iucci09}.
 
When the systems are far away from integrability, for example in higher dimensions, the expectation 
is that they should thermalize. Recent numerical work has shown this to be the case for bosons in 
two dimensions, studied with exact diagonalization \cite{rigol08STATc}, and for fermions studied 
within the dynamical mean-field theory (DMFT) approximation \cite{eckstein09}. In addition, 
thermalization in the isolated two-dimensional bosonic case could be 
understood within the eigenstate thermalization hypothesis (ETH) proposed by Deutsch \cite{deutsch91} 
and Srednicki \cite{srednicki94}, in which the individual eigenstates of the many-body Hamiltonian 
exhibit thermal behavior \cite{rigol08STATc}. 

In one-dimensional nonintegrable systems, the situation is a bit more subtle as many models can be 
tuned (by changing some Hamiltonian parameters) to be arbitrarily close to integrable points. 
An early study of the relaxation dynamics in fermionic systems when breaking integrability, using 
time-dependent density renormalization group (tDMRG) \cite{manmana07}, concluded that 
thermalization does not occur even if the system is perturbed away from the integrable point. Another 
early work in 1D, also using tDMRG, studied the relaxation dynamics of the (nonintegrable) 
Bose-Hubbard model when quenching the system across the superfluid-to-Mott-insulator transition 
\cite{kollath07}. In that case the authors concluded that thermalization occurs in some regimes but 
not in others.

More recently, we have performed a systematic study of thermalization in finite 1D systems of 
hardcore bosons. Our results indicate that far away from integrability the system does thermalize 
\cite{rigol09STATa}. However, thermalization breaks down as one approaches the integrable point. An 
important question that still needs to be answered is what happens with the point at which 
thermalization breaks down as one approaches the thermodynamic limit. It may either move toward 
the integrable point or toward some point away from integrability. More powerful numerical 
techniques or analytical approaches may be required to answer that question, which will not be 
addressed here.
 
In this work, we extend the analysis in Ref.\ \cite{rigol09STATa} to the fermionic case, which 
allows us to discuss some of the issues that remained open in Ref.\ \cite{manmana07}. In 
particular, the time scales required for different observables to relax to a stationary 
``equilibrium'' distribution, and the description of the observables after relaxation.
We also study how the closeness to an integrable point affects the dynamics and thermalization 
after a quantum quench. As in earlier works \cite{rigol08STATc,rigol09STATa}, we show that the 
breakdown of thermalization for our observables is directly linked to the breakdown of ETH.

The presentation is organized as follows. In Sec.\ \ref{Sec:hamiltF}, we introduce the
lattice model to describe 1D fermions with nearest- and next-nearest-neighbor hopping 
and repulsive interactions. We define our observables of interest and briefly discuss 
our numerical approach. The nonequilibrium dynamics of this model is presented in 
Sec.\ \ref{Sec:dynamicsF}. Section \ref{Sec:thermodynamicsF} is devoted to the analysis of
statistical mechanical approaches to describe observables after relaxation.
In Sec.\ \ref{Sec:ETHF}, we justify these statistical mechanical approaches, or their failure,
within ETH. An explanation of why the momentum distribution function of the fermions has a
distinctively slow relaxation dynamics is presented in Sec.\ \ref{Sec:OffDiagonal} in terms
of the off-diagonal elements of this observable in the basis of the eigenstates of the 
Hamiltonian. Finally, the summary and outlook are presented in Sec.\ \ref{Sec:summary}.

\section{Hamiltonian and observables}
\label{Sec:hamiltF}

The Hamiltonian of spin-polarized fermions in a 1D lattice with periodic boundary conditions can
be written as
{\setlength\arraycolsep{0.5pt}
\begin{eqnarray}
\hat{H}&=&\sum_{i=1}^L \left\lbrace -t\left( \hat{f}^\dagger_i \hat{f}_{i+1} + \textrm{H.c.} \right) 
+V\left( \hat{n}_i-\dfrac{1}{2}\right)\left( \hat{n}_{i+1}-\dfrac{1}{2}\right) \right.\nonumber \\
&-&\left.t'\left( \hat{f}^\dagger_i \hat{f}_{i+2} + \textrm{H.c.} \right)  
+V'\left( \hat{n}_i-\dfrac{1}{2}\right)\left( \hat{n}_{i+2}-\dfrac{1}{2}\right)\right\rbrace. 
\label{Eq:hamiltonianF}
\end{eqnarray}
}where the fermionic creation and annihilation operators at site 
$i$ are denoted by $\hat{f}^{\dagger}_{i}$ and $\hat{f}^{}_{i}$, respectively, 
and the local-density operator by $\hat{n}_i=\hat{f}^{\dagger}_{i}\hat{f}^{}_{i}$.
In Eq.\ (\ref{Eq:hamiltonianF}), the nearest- and next-nearest-neighbor hopping parameters 
are denoted by $t$ and $t'$, respectively, and the nearest- and next-nearest-neighbor 
interactions are denoted by $V$ and $V'$, respectively. In our study, we only consider 
repulsive interactions ($V,V'>0$), and the number of lattice sites is denoted by $L$.

In Ref.\ \cite{rigol09STATa}, we have already studied a very similar Hamiltonian
for hardcore bosons (bosons with an infinite on-site repulsion), in which case
{\setlength\arraycolsep{0.5pt}
\begin{eqnarray}
\hat{H}_{b}&=&\sum_{i=1}^L \left\lbrace -t\left( \hat{b}^\dagger_i \hat{b}_{i+1} + \textrm{H.c.} \right) 
+V\left( \hat{n}^b_i-\dfrac{1}{2}\right)\left( \hat{n}^b_{i+1}-\dfrac{1}{2}\right) \right.\nonumber \\
&-&\left.t'\left( \hat{b}^\dagger_i \hat{b}_{i+2} + \textrm{H.c.} \right)  
+V'\left( \hat{n}^b_i-
\dfrac{1}{2}\right)\left( \hat{n}^b_{i+2}-\dfrac{1}{2}\right)\right\rbrace. 
\label{Eq:hamiltonianHCB}
\end{eqnarray}
}where the hardcore boson creation and annihilation operators at site 
$i$ are denoted by $\hat{b}^{\dagger}_{i}$ and $\hat{b}^{}_{i}$, respectively, 
and the local-density operator by $\hat{n}^b_i=\hat{b}^{\dagger}_{i}\hat{b}^{}_{i}$.
The parameters $t,\,V,\,t',$ and $V'$ in Eq.\ (\ref{Eq:hamiltonianHCB}) have exactly the same meaning 
for hardcore bosons as for the fermions in Eq.\ (\ref{Eq:hamiltonianF}). For hardcore 
bosons, the creation and annihilation operators commute as usual for bosons
\begin{equation}
[\hat{b}^{}_{i},\hat{b}^{\dagger}_{j}]=
[\hat{b}^{}_{i},\hat{b}^{}_{j}]=
[\hat{b}^{\dagger}_{i},\hat{b}^{\dagger}_{j}]=0,\quad \mathrm{for} \quad i\neq j,
\end{equation}
but on the same site, the hardcore bosons operators satisfy anticommutation 
relations typical of fermions,
\begin{equation}  
\left\lbrace  \hat{b}^{}_{i},\hat{b}^{\dagger}_{i}\right\rbrace =1, 
\qquad
\hat{b}^{\dagger 2}_{i}= \hat{b}^2_{i}=0.
\label{ConstHCB} 
\end{equation}
These constraints avoid double or higher occupancy of the lattice sites. 

Similar to the hardcore boson case studied in Ref.\ \cite{rigol09STATa}, the
Hamiltonian (\ref{Eq:hamiltonianF}) is integrable for $t'=V'=0$. In this section,
we restrict our analysis to systems with 1/3 filling, and study the effects 
of finite, but small, values of $t'=V'\neq0$ when one departs from the integrable 
point. In this case, the ground state of our Hamiltonian is always a 
Luttinger liquid and no metal-insulator transition occurs in the system 
\cite{zhuravlev97}.

For $t'=0$, the fermionic Hamiltonian (\ref{Eq:hamiltonianF}) can be mapped onto the 
identical Hamiltonian for hardcore bosons (\ref{Eq:hamiltonianHCB}) (up to a boundary 
effect), i.e., the fermionic operators in Eq.\ (\ref{Eq:hamiltonianF}) just need to 
be replaced by operators describing hardcore bosons. This can be done 
using the Jordan-Wigner transformation \cite{jordan28}, and implies that in the 
thermodynamic limit both systems 
have the same spectrum and identical diagonal correlations. However, off-diagonal 
correlations and related observables, such as the momentum distribution function, are 
very different for fermions and bosons. A finite value of $t'$ still allows for a 
mapping of the fermionic Hamiltonian (\ref{Eq:hamiltonianF}) onto a Hamiltonian of
hardcore bosons. However, the new hardcore boson Hamiltonian for $t'\neq 0$ will be
different to the one in Eq.\ (\ref{Eq:hamiltonianHCB}). The mapping introduces 
additional operators in the term proportional to $t'$. In addition, in the fermionic 
case, a finite value of $t'$ breaks the particle-hole symmetry present for $t'=0$ 
and present for any value of $t'$ in the hardcore boson case.

The study of the nonequilibrium dynamics and thermodynamics of these systems is
performed using full exact diagonalization of the Hamiltonian (\ref{Eq:hamiltonianF}). 
For our largest system sizes, eight fermions in 24 sites, the total Hilbert space 
has dimension $D=735\,471$. We take advantage of the translational symmetry of 
the lattice, which allows us to block diagonalize the full Hamiltonian while the
dimension of the largest block (in $k$ space) is $D_k=30\,667$.

In this work, we focus our analysis on two observables. The first one in the 
momentum distribution function
\begin{equation}
 \hat{n}(k)=\dfrac{1}{L}\sum_{i,j} e^{-k(i-j)} \hat{f}^{\dagger}_i\hat{f}^{}_j,
\end{equation}
which is the Fourier transform of the one-body density matrix 
($\hat{\rho}_{ij}=\hat{f}^{\dagger}_i\hat{f}^{}_j$). As mentioned before, this observable
behaves very differently for fermions as compared with the hardcore bosons to which they can
be mapped. We should add that $n(k)$ is usually measured in experiments 
with ultracold quantum gases via time-of-flight expansion.

The second observable of interest is the density-density structure factor
\begin{equation}
\hat{N}(k)=\dfrac{1}{L}\sum_{i,j} e^{-k(i-j)} 
\hat{n}_i\hat{n}_j,
\end{equation}
which is the Fourier transform of the density-density correlation matrix. Since we work
at fixed number of fermions $N_f$, the expectation value of $\hat{N}(k=0)$ is always
$\langle \hat{N}(k=0)\rangle=N_f^2/L$ and, as usual, we set it to zero by subtracting
that trivial constant. The structure factor can be measured in cold gases experiments 
by means of noise correlations \cite{altman04}.

\section{Nonequilibrium dynamics}
\label{Sec:dynamicsF}

We will restrict our study of the nonequilibrium dynamics to the case in which 
the system is taken out of equilibrium by means of a quench. Quenches are a special 
way to induce dynamics by starting with an eigenstate of some initial 
Hamiltonian $\hat{H}_{ini}$, and then instantaneously changing the Hamiltonian 
(at time $\tau=0$) to some final time-independent Hamiltonian $\hat{H}_{fin}$. 

As mentioned previously, our model is integrable for $t'=V'=0$. Since we are interested in 
studying the effect that approaching an integrable point has on the dynamics and 
thermalization of the system, we take the initial state to be an eigenstate of a system 
with $t=t_{ini}$, $V=V_{ini}$, $t'$, $V'$, and then quench the nearest-neighbor parameters 
$t$ and $V$ to $t=t_{fin}$, $V=V_{fin}$ without changing $t'$, $V'$, i.e., we only 
change $t_{ini},V_{ini}\rightarrow t_{fin},V_{fin}$. This quench is performed
for different values of $t'$, $V'$ as one approaches the integrable point $t'=V'=0$.

Notice that our quenches do not break translational symmetry. For that reason, we always select 
the initial state to be one of the eigenstates of the initial Hamiltonian in the total $k=0$ 
sector. Since translational symmetry is preserved, only states with zero total momentum are 
required to calculate the exact time evolution of the system
\begin{equation}
\WFtime=e^{-i\widehat{H}_{fin}\tau}\WFini= 
\sum_\alpha C_{\alpha}e^{-i E_\alpha \tau}\EigenvO \, ,
\end{equation}
where $\WFtime$ is the time-evolving state, $\WFini$ is the initial state, 
$\EigenvO$ are the eigenstates of the final Hamiltonian with zero total momentum, 
energy $E_\alpha$, and $C_{\alpha}=\langle \Psi^0_\alpha\WFini$. The sum runs
over all the eigenstates of the total $k=0$ sector. In this section, we perform the exact 
time evolution of up to eight fermions in lattices with up to 24 sites. This means 
that the largest total $k=0$ sector diagonalized had 30\,666 states.

In contrast to classical systems, where one has to perform the time evolution in order 
to compute the long-time average of any observable, in isolated quantum systems (where 
the time evolution is unitary) one can predict such time averages without the need of calculating
the dynamics. Since the wave function of any initial state can be written in terms of the 
eigenstates of the final Hamiltonian $\Eigenvleft$ as $\WFini=\sum_\alpha C_{\alpha}\Eigenvleft$, 
one finds that the time evolution of the quantum expectation value of an observable $\hat{O}$ 
can be written as
\begin{equation}
\langle \widehat{O}(\tau) \rangle
\equiv\langle \psi(\tau) | \widehat{O} | \psi(\tau)  \rangle
=\sum_{\leftindex,\,\rightindex} C_{\leftindex}^{\star} C_{\rightindex}^{}
e^{i(E_{\leftindex}-E_{\rightindex})\tau}
O_{\leftindex\rightindex}\, ,
\label{Eq:timeevolution}
\end{equation}
where $O_{\alpha\beta}$ are the matrix elements of $\hat{O}$ in the basis of the final 
Hamiltonian. This in turn implies that the infinite time average of the observable can be written as
\begin{eqnarray}
\overline{\langle \widehat{O}(\tau) \rangle} \equiv O_{diag}=\sum_{\alpha} |C_{\alpha}|^{2}
O_{\alpha\alpha}.
\label{Eq:diagonal}
\end{eqnarray}
We have assumed that, as in the case of generic (nonintegrable) systems, the spectrum 
is nondegenerate and incommensurate. This means that if the expectation value of $\hat{O}$ 
relaxes to some kind of equilibrium value (up to the recurrences that must occur if the system
is isolated), that value should be the one predicted by Eq.\ (\ref{Eq:diagonal}). Following 
previous work, we prefer to think of this exact result as the prediction of a 
``diagonal ensemble,'' where $|C_{\alpha}|^{2}$ is the weight of each state within this 
ensemble \cite{rigol08STATc,rigol09STATa}.

One objection that may arise at this point is that even if the system is nondegenate and
incommensurate, the spectrum may have arbitrarily close levels and it may take an unrealistically 
long time for the prediction in Eq.\ (\ref{Eq:diagonal}) to apply, i.e., that it may not be relevant 
to describe experiments. In order to study how the dynamics drives our observables of interest 
[$n(k)$ and $N(k)$] toward the prediction of Eq.\ (\ref{Eq:diagonal}), we follow the scheme in 
Ref.\ \cite{rigol09STATa}. We study the normalized area between $n(k,\tau)$ and $N(k,\tau)$, 
during the time evolution, and their infinite time average, i.e., at different times we compute
\begin{equation}
 \delta n_k(\tau)=\dfrac{\sum_k|n(k,\tau)-n_{diag}(k)|}{\sum_k n_{diag}(k)}
\label{Eq:errorn}
\end{equation}
and
\begin{equation}
 \delta N_k(\tau)=\dfrac{\sum_k|N(k,\tau)-N_{diag}(k)|}{\sum_k N_{diag}(k)}
\label{Eq:errorN}
\end{equation}

\begin{figure}[!h]
\begin{center}
\includegraphics[width=0.48\textwidth,angle=0]{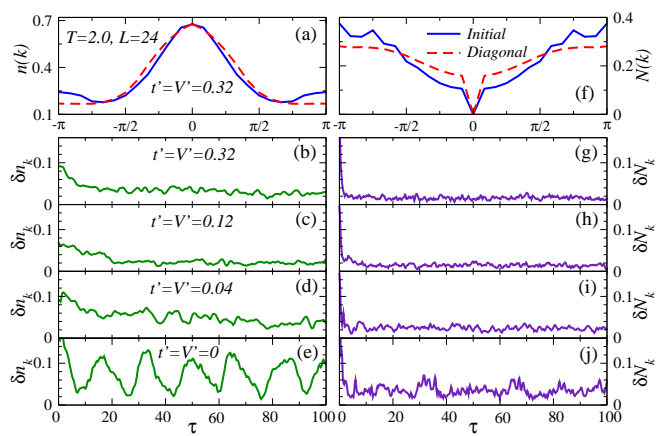}
\end{center}
\vspace{-0.6cm}
\caption{\label{Fig:TimeEvolution_L24T2.0} (Color online)
Quantum quench $t_{ini}=0.5$, $V_{ini}=2.0$ $\rightarrow$ $t_{fin}=1.0$, $V_{fin}=1.0$, 
with $t'_{ini}=t'_{fin}=t'$ and $V'_{ini}=V'_{fin}=V'$ in a system with $N_f=8$ and 
$L=24$. The initial state was chosen in such a way that after the quench the system 
has an effective temperature $T=2.0$ \cite{temperature} in all cases. 
Given the energy of the initial state in the final Hamiltonian 
$E=\langle\psi_{ini}\vert \widehat{H}_{fin}\vert \psi_{ini}\rangle$, 
the effective temperature is computed following Eq.\ (\ref{Eq:Temperature}). 
(a) Initial and diagonal ensemble results for $n(k)$ when $t'=V'=0.32$. 
(b)--(e) Time evolution of $\delta n_k$ for $t'=V'=0.32$, 0.12, 0.04, and 0.0, respectively. 
(f) Initial and diagonal ensemble results for $N(k)$ when $t'=V'=0.32$. 
(g)--(j) Time evolution of $\delta N_k$ for $t'=V'=0.32$, 0.12, 0.04, and 0.0, respectively.}
\end{figure}

In Fig.\ \ref{Fig:TimeEvolution_L24T2.0}, we show results for $\delta n_k(\tau)$ and $\delta N_k(\tau)$ 
as a function of time $\tau$ for four different quenches as one approaches the 
integrable point. Besides undergoing the same change $t_{ini}=0.5$, $V_{ini}=2.0$ 
$\rightarrow$ $t_{fin}=1.0$, $V_{fin}=1.0$, these quenches have another very important 
property in common. The initial state for each was selected to be one  
eigenstate of the initial Hamiltonian in such a way that the effective temperature 
$T$ of the system \cite{temperature} is always the same ($T=2.0$ in 
Fig.\ \ref{Fig:TimeEvolution_L24T2.0}). Given the energy of the time-evolving state 
in the final Hamiltonian, which is conserved, 
\begin{equation}
 E=\langle\psi_{ini}\vert \widehat{H}_{fin}\vert \psi_{ini}\rangle, 
\end{equation}
the effective temperature $T$ \cite{temperature} is defined by the expression
\begin{equation}
 E=\dfrac{1}{Z}\textrm{Tr}\left\lbrace \hat{H}_{fin} e^{-\hat{H}_{fin}/{T}}\right\rbrace, 
\label{Eq:Temperature}
\end{equation}
where
\begin{equation}
 Z=\textrm{Tr}\left\lbrace e^{-\hat{H}_{fin}/{T}}\right\rbrace
\end{equation} 
is the partition function, and we have set the Boltzmann constant $k_B$ to unity, and 
$t_{fin}=1$ sets the energy scale in our system.

In Figs.\ \ref{Fig:TimeEvolution_L24T2.0}(a) and \ref{Fig:TimeEvolution_L24T2.0}(f), we 
compare the initial momentum distribution and structure factors with the predictions of 
Eq.\ (\ref{Eq:diagonal}) away from integrability ($t'=V'=0.32$). (The results for other 
values of $t',V'$ are similar and not shown here.) The dynamics of those observables for 
four different values of $t',V'$ is presented in 
Figs.\ \ref{Fig:TimeEvolution_L24T2.0}(b)--\ref{Fig:TimeEvolution_L24T2.0}(e) for $\delta n_k(\tau)$ and in 
Figs.\ \ref{Fig:TimeEvolution_L24T2.0}(g)--\ref{Fig:TimeEvolution_L24T2.0}(j) for $\delta N_k(\tau)$. They
show that the dynamics of $n(k)$ and $N(k)$ are different. Away from integrability 
($t'=V'\neq0$), $\delta N_k(\tau)$ quickly relaxes [in a time scale $\tau\sim (t_{fin})^{-1}=1.0$] to a value 
$\sim\,$0.02 and then fluctuates around that value. $\delta n_k(\tau)$, on the other hand, slowly drifts toward
$\delta n_k(\tau)\,$$\sim\,$$0.02$--$0.04$ in a much longer time scale, orders of magnitude longer than the 
one seen for the relaxation of $\delta N_k(\tau)$.

The behavior of $\delta N_k(\tau)$ in these spin-polarized systems is 
qualitatively (and quantitatively) very similar to the one observed for 
hardcore bosons in Ref.\ \cite{rigol09STATa}. The time evolution of 
$\delta n_k$ for the fermions, on the other hand, is different from the one 
seen for the hardcore bosons in Ref.\ \cite{rigol09STATa}. For the latter 
systems, $n(k)$ quickly relaxed toward the predictions of the diagonal ensemble,
even at integrability. In Fig.\ \ref{Fig:TimeEvolution_L24T2.0}(j), one can see 
that at integrability $\delta N_k(\tau)$ for the fermions also relaxes quickly toward 
the predictions of the diagonal ensemble and then exhibits fluctuations 
between $\delta N_k(\tau)=0.02$ and $\delta N_k(\tau)=0.06$, while $\delta n_k(\tau)$ in 
Fig.\ \ref{Fig:TimeEvolution_L24T2.0}(e) exhibits very large oscillations between 
$\delta n_k(\tau)=0.02$ and $\delta n_k(\tau)=0.12$. Additionally, we studied the dynamics
of $\delta n_k$ for a time scale ten times longer than the one depicted in 
Fig.\ \ref{Fig:TimeEvolution_L24T2.0}(e) and observed exactly the same behavior.

\begin{figure}[!h]
\begin{center}
\includegraphics[width=0.48\textwidth,angle=0]{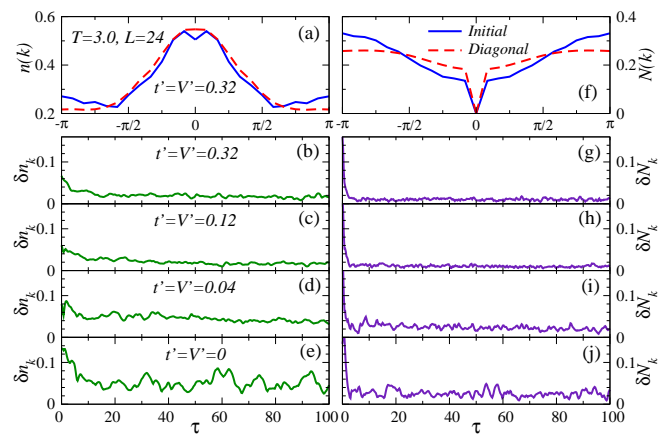}
\end{center}
\vspace{-0.6cm}
\caption{\label{Fig:TimeEvolution_L24T3.0}  (Color online)
Quantum quench $t_{ini}=0.5$, $V_{ini}=2.0$ $\rightarrow$ $t_{fin}=1.0$, $V_{fin}=1.0$, 
with $t'_{ini}=t'_{fin}=t'$ and $V'_{ini}=V'_{fin}=V'$ in a system with $N_f=8$ 
and $L=24$. The initial state was chosen in such a way that after the quench the 
system has an effective temperature $T=3.0$ \cite{temperature} in all cases 
(see the caption of Fig.\ \ref{Fig:TimeEvolution_L24T2.0}). (a) Initial and 
diagonal ensemble results for $n(k)$ and $t'=V'=0.32$. (b)--(e) Time evolution 
of $\delta n_k$ for $t'=V'=0.32$, 0.12, 0.04, and 0.0, respectively. 
(f) Initial and diagonal ensemble results for $N(k)$ and $t'=V'=0.32$. 
(g)--(j) Time evolution of $\delta N_k$ for $t'=V'=0.32$, 0.12, 0.04, 
and 0.0, respectively.}
\end{figure}

For hardcore bosons \cite{rigol09STATa}, we have shown that increasing the effective 
temperature \cite{temperature} (or the final energy $E$) decreases the mean value of 
$\delta n_k(\tau)$ and $\delta N_k(\tau)$ after relaxation and also decreases the 
temporal fluctuations of those quantities. In Fig.\ \ref{Fig:TimeEvolution_L24T3.0}, 
we show results for quenches similar to the ones in Fig.\ \ref{Fig:TimeEvolution_L24T2.0}, 
but with a higher effective temperature ($T=3.0$). The comparison between 
Figs.\ \ref{Fig:TimeEvolution_L24T3.0} and \ref{Fig:TimeEvolution_L24T2.0} 
also reveals a reduction in the mean values of $\delta n_k(\tau)$ and $\delta N_k(\tau)$ and their 
fluctuations, which is particularly obvious at integrability. Still, differences remain between
the relaxation dynamics of $\delta n_k(\tau)$ and $\delta N_k(\tau)$. Also, one can notice that at 
integrability the mean values of $\delta n_k(\tau)$ and $\delta N_k(\tau)$ and their fluctuations 
are in general larger than away from integrability, and they all decrease the further away 
one moves from the integrable point.

\begin{figure}[!h]
\begin{center}
\includegraphics[width=0.48\textwidth,angle=0]{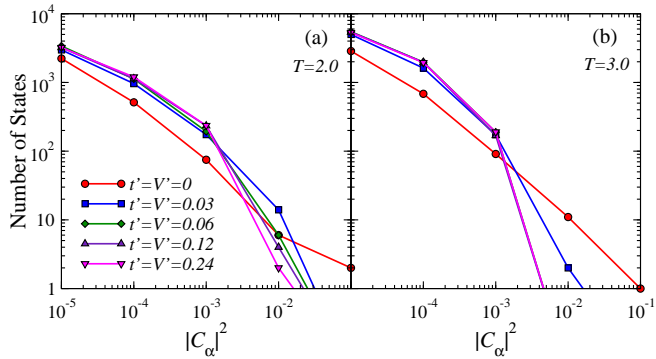}
\end{center}
\vspace{-0.6cm}
\caption{\label{Fig:StateCount}  (Color online)
Number of states with $|C_\alpha|^2$ greater than the value presented in the 
$x$ axis, for an effective temperature (a) $T=2.0$ and (b) $T=3.0$, 
and for the same quenches studied in Figs.\ \ref{Fig:TimeEvolution_L24T2.0} 
and \ref{Fig:TimeEvolution_L24T3.0}, respectively. Here, $N_f=8$ and $L=24$.}
\end{figure}

The improvement of the relaxation dynamics with increasing the effective temperature 
(energy) of the isolated state \cite{temperature}, discussed above, 
can be related in part to the increase 
in the density of states in the final system with increasing energy. This increases the 
number of eigenstates of the Hamiltonian that participate in the dynamics making dephasing
in Eq.\ (\ref{Eq:timeevolution}) more effective, and hence reducing temporal fluctuations after
relaxation. This can be better seen in Fig.\ \ref{Fig:StateCount}, which clearly shows that the number 
of states with the largest values of $|C_\alpha|^2$ increases as the temperature decreases. Since
$\sum_\alpha|C_\alpha|^2=1$, this means that the lower the temperature the smaller the total
number of states that participate in the dynamics. This can be also seen in Fig.\ \ref{Fig:StateCount},
where the number of states with $|C_\alpha|^2>10^{-5}$ and $|C_\alpha|^2>10^{-4}$ is larger for 
$T=3.0$ [Fig.\ \ref{Fig:StateCount}(b)] than for $T=2.0$ [Fig.\ \ref{Fig:StateCount}(a)] 
(notice the logarithmic scale in both axes). Overall, one can conclude that dephasing becomes 
less effective with decreasing temperature and this leads to an enhancement of temporal fluctuations.

A similar argument, based on the number of states participating in the dynamics, 
can help us understand why the average values and temporal fluctuations of $\delta n_k(\tau)$ 
and $\delta N_k(\tau)$ are in general larger after relaxation when one is at integrability, 
or close to it. Figure \ref{Fig:StateCount} also shows that for any given temperature,
the total number of states that participate in the dynamics decreases as one 
approaches the integrable point. At integrability, this may be related to the presence
of additional conserved quantities, which restrict the number of eigenstates of the 
final Hamiltonian that can have a significant overlap with the initial state.

\begin{figure}[!h]
\begin{center}
\includegraphics[width=0.48\textwidth,angle=0]{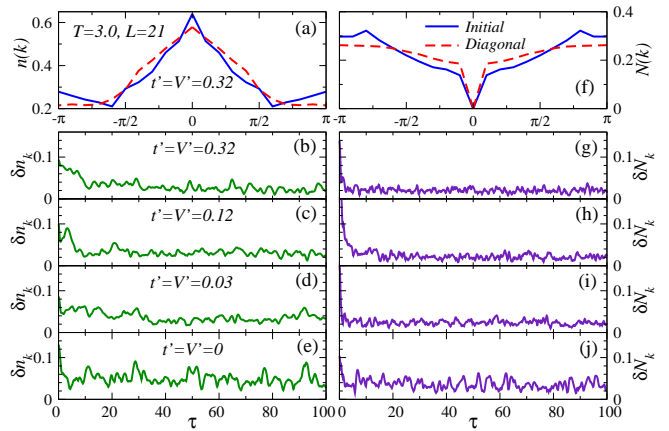}
\end{center}
\vspace{-0.6cm}
\caption{\label{Fig:TimeEvolution_L21T3.0} (Color online)
Quantum quench $t_{ini}=0.5$, $V_{ini}=2.0$ $\rightarrow$ $t_{fin}=1.0$, $V_{fin}=1.0$, 
with $t'_{ini}=t'_{fin}=t'$ and $V'_{ini}=V'_{fin}=V'$ in a system with $N_f=7$ 
and $L=21$. The initial states were selected in the same way as the ones in 
Fig.\ \ref{Fig:TimeEvolution_L24T3.0} ($T=3.0$). (a) Initial and diagonal ensemble 
results for $n(k)$ and $t'=V'=0.32$. (b)--(e) Time evolution of 
$\delta n_k$ for $t'=V'=0.32$, 0.12, 0.04, and 0.0, respectively. 
(f) Initial and diagonal ensemble results for $N(k)$ and $t'=V'=0.32$. 
(g)--(j) Time evolution of $\delta N_k$ for $t'=V'=0.32$, 0.12, 0.04, 
and 0.0, respectively.}
\end{figure}

It would be desirable to study how the mean values and temporal fluctuations of 
$\delta n_k(\tau)$ and $\delta N_k(\tau)$ behave after relaxation as one increases the system size.
Unfortunately, the computational requirements of our approach
increase exponentially with system size and a rigorous finite-size scaling is not possible.
As a step in understanding how finite-size effects affect our results here, 
we show in Fig.\ \ref{Fig:TimeEvolution_L21T3.0} results for a smaller system with
seven fermions in 21 lattice sites (with $T=3.0$). By comparing
those results with the ones in Fig.\ \ref{Fig:TimeEvolution_L24T3.0} one can see that, as
expected, decreasing the system size increases the mean values of $\delta n_k(\tau)$ and 
$\delta N_k(\tau)$ after relaxation and the temporal fluctuations of both quantities. This 
supports the expectation that with increasing the system size, after relaxation, 
the mean values of $\delta n_k(\tau)$ and $\delta N_k(\tau)$ should become arbitrarily small 
and so should the temporal fluctuations.

A different question is what happens to the time scale for $n(k)$ to relax to 
the diagonal ensemble prediction as the system size increases. Previous work
on that direction \cite{moeckel08,moeckel09} has suggested a possible intermediate 
quasisteady regime that occurs before full relaxation. This has been seen in
recent numerical work \cite{eckstein09}, and our results for the behavior of 
$n(k)$ in smaller systems may be an indication in that direction. Something that is 
important to keep in mind from our results in this section is that different observables 
may exhibit different relaxation times. In particular, and in contrast to the hardcore 
boson case, the momentum distribution function of fermions is an observable that 
takes a long time to relax to an equilibrium distribution. This may have
influenced the failure to observe thermalization in the momentum distribution function 
of a nonintegrable fermionic system in Ref.\ \cite{manmana07}.

It is interesting to note that if $V=t'=V'=0$ (noninteracting case) the 
momentum distribution function of fermions in a periodic system is a conserved 
quantity, i.e., for any initial state it will not change in time. However, the 
$n(k)$ of hardcore bosons, to which the fermions can be mapped, can have a nontrivial
dynamics and relaxes to the predictions of a generalized Gibbs ensemble, as 
shown in Refs.\ \cite{rigol07STATa} and \cite{rigol06STATb}. Having $V,V'\neq0$ (in our case 
$V=t=1.0$ and $V'$ varies between 0 and 0.32) allows the $n(k)$ of fermions to change with 
time (since now the fermions are interacting), but the time scale for relaxation to the equilibrium 
distribution still seems to be affected by the fact that in $k$ space the scattering 
between fermions is very special. We find the relaxation time for $n(k)$ to be much 
longer than the corresponding time scale for other observables such as $N(k)$. The 
differences between the time scales for the relaxation of different observables 
(and their temporal fluctuations) can be understood in terms of the off-diagonal matrix elements 
of the observables, which will be discussed in Sec.\ \ref{Sec:OffDiagonal}.

\section{Thermodynamics}
\label{Sec:thermodynamicsF}

In Sec.\ \ref{Sec:dynamicsF}, we have shown that, for our systems of interest,
the diagonal ensemble provides an accurate description of observables after
relaxation. Some observables may take longer to relax, and under some circumstances
they may not even relax at all [see, {\it e.g.}, $n(k)$ in 
Fig.\ \ref{Fig:TimeEvolution_L24T2.0}(e)], but whenever we find relaxation to an
equilibrium value, it is well described by the predictions of Eq.\ (\ref{Eq:diagonal}).
This was expected and we refer the reader to 
Refs.\ \cite{rigol08STATc,kollar08,rigol09STATa} for similar results in other 
quantum systems. 
In what follows, we study how the predictions of the diagonal ensemble compare to 
standard statistical mechanical ensembles. A great advantage of the infinite time 
average [Eq.\ (\ref{Eq:diagonal})] is that all time dependence has been removed from 
the time evolution of the quantum-mechanical problem and one has a unique 
result to test statistical mechanics.

Since the systems on which we have performed the dynamics are isolated, the most
appropriate statistical ensemble to compare observables after relaxation is the 
microcanonical ensemble. As usual, the computations in the microcanonical ensemble 
are performed averaging over all eigenstates (from all momentum sectors) that lie 
within a window $[E-\Delta E, E+\Delta E]$, where 
$E=\langle \psi_{ini} | \hat{H}_{fin} | \psi_{ini}\rangle$, and we have 
taken $\Delta E=0.1$ in all cases. Similarly to what was done in 
Refs.\ \cite{rigol08STATc} and \cite{rigol09STATa}, we have checked that our results are 
independent of the exact value of $\Delta E$ in the neighborhood of $\Delta E=0.1$.

\begin{figure}[!h]
\begin{center}
\includegraphics[width=0.36\textwidth,angle=0]{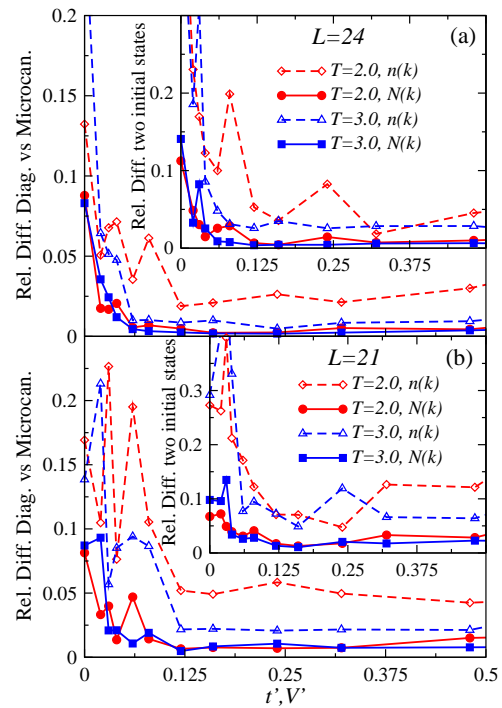}
\end{center}
\vspace{-0.6cm}
\caption{\label{Fig:Thermodynamics} (Color online)
(Main panels) Comparison between the prediction of the microcanonical
and diagonal ensembles, for $n(k)$ and $N(k)$, as a function of increasing
$t',V'$ for $T=2.0$ and $T=3.0$. Results are shown for (a) $L=24$, $N_f=8$, 
and (b) $L=21$, $N_f=7$. The diagonal ensembles correspond to the quenches
in Figs.\ \ref{Fig:TimeEvolution_L24T2.0}, \ref{Fig:TimeEvolution_L24T3.0},
and \ref{Fig:TimeEvolution_L21T3.0}.
(Insets) Comparison between the prediction of diagonal ensembles generated by two
different initial states, for $T=2.0$ and $T=3.0$. The results shown are for 
(a) $L=24$, $N_f=8$, and (b) $L=21$, $N_f=7$. As in the main panels, one of the 
diagonal ensembles is generated by initial states selected from the eigenstates 
of a Hamiltonian with $t_{ini}=0.5$, $V_{ini}=2.0$, the other diagonal ensemble 
is generated by initial states selected from the eigenstates of a Hamiltonian with 
$t_{ini}=2.0$, $V_{ini}=0.5$. The final Hamiltonian (with $t_{fin}=1.0$, 
$V_{fin}=1.0$) and the effective temperature \cite{temperature} are identical 
for both initial states. By relative differences in this figure we mean the 
normalized area between the different ensemble predictions for $n(k)$ and 
$N(k)$. Relative differences are computed in exactly the same way as 
$\delta n_k(\tau)$ and $\delta N_k(\tau)$ are computed in Eqs.\ (\ref{Eq:errorn}) 
and (\ref{Eq:errorN}), respectively.}
\end{figure}

The main panel in Fig.\ \ref{Fig:Thermodynamics}(a) depicts how the difference between
the diagonal and the microcanonical ensembles behaves as one moves away from 
integrability. Once again, we find a different behavior for $n(k)$ and $N(k)$. 
For $N(k)$, we find that for both effective temperatures \cite{temperature} 
considered, the difference between the diagonal and microcanonical ensemble 
is always smaller than 1\% for $t'=V'>0.06$, and one can say that the system 
exhibits thermal behavior. This is very similar to the results for the same 
quantity obtained in the hardcore boson systems analyzed in 
Ref.\ \cite{rigol09STATa}. On the other hand, $n(k)$ 
exhibits a larger difference between the diagonal and the microcanonical ensemble
for the same values of $t',V'$, in particular at lower temperatures. For both 
observables, one can still see that as one approaches the integrable point the 
difference between both ensembles increases, signaling the breakdown of thermalization 
in all cases.

In the inset in Fig.\ \ref{Fig:Thermodynamics}(a), we also compare the predictions 
of two diagonal ensembles generated by different initial states, which are chosen 
in such a way that the effective temperature of both system after the quench 
\cite{temperature} are the same. The behavior is qualitatively similar to the 
one seen in the main panel of the same figure and shows that the breakdown of 
thermalization for both observables is accompanied by a dependence on the initial 
state.

Results for a smaller system size, with the same density and effective temperatures,
are shown in the main panel and inset in Fig.\ \ref{Fig:Thermodynamics}(b). They show 
that, as expected, the differences between the two ensembles for any given value of
$t',V'$ increase with decreasing system size. Notice that the $y$ axis in 
Figs.\ \ref{Fig:Thermodynamics}(a) and \ref{Fig:Thermodynamics}(b) has a 
different scale, so that the contrast between the results in both panels is stronger 
than may appear at first sight. 

For $t'=V'>0.06$, the comparison between the results for different systems sizes led us
to expect that the predictions of the microcanonical ensemble, for both observables, will
coincide with the ones of the diagonal ensemble as the system size is increased, i.e.,
that thermalization takes place if one is sufficiently far from integrability. We do note 
that $n(k)$ for these fermionic systems is more sensitive to finite-size effects
than $N(k)$, something that is in contrast to the behavior seen for hardcore bosons 
\cite{rigol09STATa}. 

For values of $t'=V'$ closer to the $t'=V'=0$ integrable point, the outcome of increasing
system size is less clear, at least for the system sizes that we are able to analyze here. Two 
possibilities emerge. (i) As the system size increases the point at which the results for the 
diagonal an microcanonical ensembles start to differ will move toward $t'=V'=0$. (ii) The
very same point will move toward a nonzero value of $t',V'$. Here, we are 
not able to discriminate between those two scenarios, and further studies will be needed to 
address that question.

\begin{figure}[!h]
\begin{center}
\includegraphics[width=0.36\textwidth,angle=0]{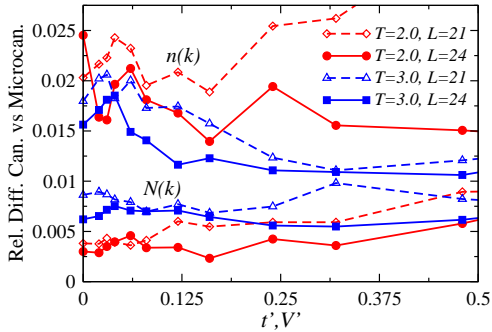}
\end{center}
\vspace{-0.6cm}
\caption{\label{Fig:Microvscanonical} (Color online)
Relative differences (normalized area) between the predictions of the 
microcanonical and canonical ensembles for $n(k)$ [upper four curves] 
and $N(k)$ [lower four curves] in systems with $L=21$, $N_f=7$, and $L=24$, 
$N_f=8$. The relative differences are computed in exactly the same way 
as $\delta n_k(\tau)$ and $\delta N_k(\tau)$ are computed in Eqs.\ (\ref{Eq:errorn}) 
and (\ref{Eq:errorN}), respectively.}
\end{figure}

It is relevant to notice that, for the small systems sizes studied in this work, it is 
important to select the microcanonical ensemble when comparing with the outcome of the dynamics.
This is because finite-size effects can make the predictions of different standard statistical 
mechanical ensembles different from each other. In Fig.\ \ref{Fig:Microvscanonical}, we compare
the results of the microcanonical ensemble, which we have been using up to this point, with the
ones obtained with the canonical ensemble. One can see there that depending on the observable 
under consideration the results of both ensembles can differ up to $\sim$ 2.5\%. The differences 
are not very much affected by the proximity to integrability, and, as expected, they can be seen to 
decrease with increasing system size. One can also see in Fig.\ \ref{Fig:Microvscanonical} that 
the differences between the two ensembles are always larger for $n(k)$. Interestingly, they are not 
as large as the ones observed for the same quantity in the case of hardcore bosons \cite{rigol09STATa}.

\section{Eigenstate thermalization hypothesis}
\label{Sec:ETHF}

In what follows, we will analyze the reason underlying thermalization in these
isolated quantum systems, and the cause for the differences seen between $n(k)$ and $N(k)$ when 
comparing the diagonal and microcanonical ensembles.

In Sec.\ \ref{Sec:thermodynamicsF}, we have shown that away from integrability there are regimes
in which the system thermalizes, i.e., in which the predictions of the diagonal ensemble and the
microcanonical ensemble coincide. This means that
\begin{eqnarray}
O_{diag}&=&O_{microc},\nonumber \\
\sum_{\alpha} |C_{\alpha}|^{2} 
O_{\alpha\alpha} &=&
\frac{1}{\mathcal{N}_{E,\, \Delta E}} \sum_{\substack{\alpha \\ 
\left|E-E_{\alpha}\right|<\Delta E}}\,O_{\alpha\alpha},
\label{Eq:paradox}
\end{eqnarray}
where $\mathcal{N}_{E,\, \Delta E}$ is the number of energy eigenstates with energies in the window 
$\left[E-\Delta E,\, E+\Delta E\right]$. This result is certainly surprising because the left-hand 
side of Eq.\ (\ref{Eq:paradox}) depends on the initial conditions through the overlaps of the 
initial state with the eigenstates of the final Hamiltonian ($C_{\alpha}$), while the right-hand side of 
Eq.\ (\ref{Eq:paradox}) only depends on the energy $E$ (as mentioned before, our results do not 
depend on the specific value of $\Delta E$). 

A possible solution to this paradox was advanced by Deutsch \cite{deutsch91} 
and Srednicki \cite{srednicki94} in terms of the eigenstate thermalization hypothesis (ETH). 
Within ETH, the difference between the eigenstate expectation values of generic few-body 
observables [$O_{\alpha\alpha}$] are presumed to be small between eigenstates that are close in 
energy. This implies that if one takes a small window of energy $\Delta E$, as it is done in the 
microcanonical ensemble, all the eigenstate expectation values within that window will be very 
similar to each other. Hence, the microcanonical average and a single eigenstate will provide essentially 
the same answer, i.e., the eigenstates already exhibit thermal behavior. The same will happen with 
the prediction of the diagonal ensemble as long as the $|C_{\alpha}|$'s are strongly peaked around 
the energy $E$. The latter has been shown to be the case for quenches in hardcore boson systems 
in two dimensions \cite{rigol08STATc}, hardcore bosons in one dimension \cite{rigol09STATa,rigol09STATb}, 
softcore bosons in one dimension \cite{roux09}, and generic quenches in the thermodynamic 
limit \cite{rigol08STATc}.

\begin{figure}[!h]
\begin{center}
\includegraphics[width=0.47\textwidth,angle=0]{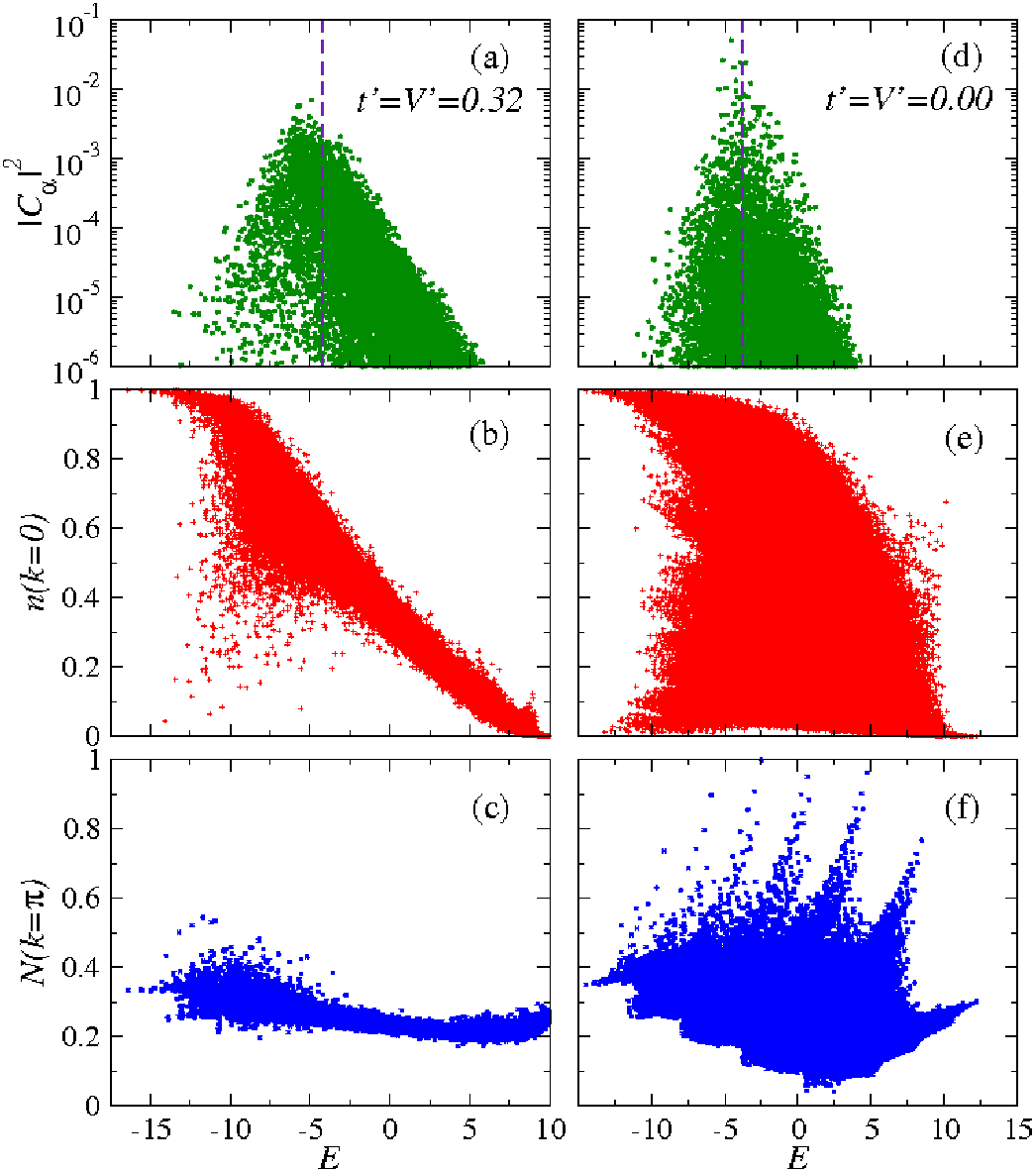}
\end{center}
\vspace{-0.6cm}
\caption{\label{Fig:ETH} (Color online)
(a),(d) Distribution of $|C_{\alpha}|^{2}$ for two of the quenches depicted in 
Fig.\ \ref{Fig:TimeEvolution_L24T3.0}, for (a) $t'_{fin}=V'_{fin}=0.32$ and
(d) $t'_{fin}=V'_{fin}=0.0$. In both cases, the final effective temperature
of the system is $T=3.0$ \cite{temperature}. The vertical dashed lines signal 
the energy $E$ of the
time-evolving state. (b),(e) Eigenstate expectation values of $\hat{n}(k=0)$ in the
full spectrum (including all momentum sectors) of the Hamiltonian (\ref{Eq:hamiltonianF}) 
with $t=V=1.0$ and, (b) $t'=V'=0.32$ and (e) $t'=V'=0.0$. (c),(f) Eigenstate 
expectation values of $\hat{N}(k=\pi)$ in the full spectrum (including all momentum 
sectors) of the Hamiltonian (\ref{Eq:hamiltonianF}) with $t=V=1.0$ and, (c) $t'=V'=0.32$ 
and (f) $t'=V'=0.0$. These results were obtained for systems with $L=24$ and $N_f=8$, i.e., 
the total Hilbert space consists of 735\,471 states.}
\end{figure}

In Figs.\ \ref{Fig:ETH}(a) and \ref{Fig:ETH}(d), we show the distributions of 
$|C_{\alpha}|^2$ for two of the quenches studied in Fig.\ \ref{Fig:TimeEvolution_L24T3.0}.
One quench is far away from integrability and the other one at integrability. There are some 
features of the distributions of $|C_{\alpha}|^2$ that are important to mention here. (i) They 
are neither similar to the microcanonical nor to the canonical distributions. (ii) They are strongly
peaked around the energy $E$ of the time-evolving state. Actually, one can see in the figures 
that the values of $|C_{\alpha}|^2$ decay exponentially as one moves away from $E$ (signaled by
the vertical line in the figures). (iii) They are not related to the effective temperature
of the system, which is exactly the same in both cases ($T=3.0$) while the distributions of 
$|C_{\alpha}|^2$ are clearly different. The specific values of $|C_{\alpha}|^2$, and the exponent 
of their decay, depend on the initial state \cite{rigol09STATb}. Because of these properties
of the $|C_{\alpha}|^2$ distributions, they alone cannot explain why thermalization
takes place, for example, for $N(k)$ when $t'=V'=0.32$ in Fig.\ \ref{Fig:Thermodynamics}.

In the bottom four panels of Fig.\ \ref{Fig:ETH}, we show the eigenstate expectation values
of $\hat{n}(k=0)$ [(b),(e)] and $\hat{N}(k=\pi)$ [(c),(f)] for a system far from integrability,
with $t=V=1.0$ and $t'=V'=0.32$ [(b),(c)], and for a system at integrability, with $t=V=1.0$ and 
$t'=V'=0.0$ [(e),(f)]. Those results, together with the $|C_{\alpha}|^2$ distributions
in Figs.\ \ref{Fig:ETH}(a) and \ref{Fig:ETH}(d), can help us understand the differences 
between the diagonal and the microcanonical ensembles in Fig.\ \ref{Fig:Thermodynamics}(a).

The most striking feature in the bottom panels in Fig.\ \ref{Fig:ETH} is the contrast between
the eigenstate expectation values at integrability and far away from it. The former ones exhibit
a very wide distribution of eigenstate expectation values. This is true even if one selects a 
narrow window at the center of the spectrum (equivalent to very high temperatures). As one moves 
away from integrability, one can clearly see 
that the fluctuations between eigenstate expectation values of nearby eigenstates reduces 
dramatically and ETH starts to be valid. For $t'=V'=0.32$, a comparison between the expectation 
values of $\hat{n}(k=0)$ and $\hat{N}(k=\pi)$ shows that for the energies of the final states created
in our quenches, the distribution of the eigenstate expectation values of $\hat{n}(k=0)$ is 
relatively broader than the one seen for $\hat{N}(k=\pi)$, which explains why the differences
between the diagonal and microcanonical ensembles in Fig.\ \ref{Fig:Thermodynamics}(a) are larger 
for the former one. It also helps in understanding why $\hat{n}(k)$ is more sensitive to the 
selection of the initial state.

In order to gain a more quantitative understanding of how ETH breaks down as one approaches 
integrability, we have computed the average relative deviation of the eigenstate expectation 
values with respect to the microcanonical prediction, $\Delta^{mic}n(k=0)$ and $\Delta^{mic}N(k=\pi)$. 
For any given energy of the microcanonical ensemble, these quantities are computed as
\begin{equation}
 \Delta^{mic}n(k=0)=\dfrac{\sum_{\alpha}\,
|n_{\alpha \alpha}(k=0)-n_{mic}(k=0)|}{\sum_{\alpha}\,n_{\alpha \alpha}(k=0)}
\end{equation}
and
\begin{equation}
\Delta^{mic}N(k=\pi)=\dfrac{\sum_{\alpha}\,
|N_{\alpha \alpha}(k=\pi)-N_{mic}(k=\pi)|}{\sum_{\alpha}\, N_{\alpha \alpha}(k=\pi)}
\end{equation}
where $n_{\alpha \alpha}(k=0)$ and $N_{\alpha \alpha}(k=\pi)$ are the eigenstate expectation values of 
$\hat{n}(k=0)$ and $\hat{N}(k=\pi)$, respectively, and $n_{mic}(k=0)$ and $N_{mic}(k=\pi)$ are the 
microcanonical expectation values at any given energy $E$. The sum over $\alpha$ contains all states 
with energies in the window $[E-\Delta E, E+\Delta E]$. As discussed before, we have taken $\Delta E=0.1$. 

Clearly, for different values of the Hamiltonian parameters, the lowest and highest-energy states as
well as the level spacing are different. Hence, in order to make a meaningful comparison between different
systems as one moves away from the integrable point, we study the behaviors of $\Delta^{mic}n(k=0)$ and 
$\Delta^{mic}N(k=\pi)$ as a function of the effective temperature $T$ of a canonical ensemble that has the
same energy $E$ as the microcanonical ensemble. Given an energy $E$ within the microcanonical 
ensemble, the effective temperature can be computed by means of Eq.\ (\ref{Eq:Temperature}). 
We should stress that the effective temperature is only used here as an auxiliary tool for 
comparing different systems, i.e., it just provides a unique energy scale for comparing observables 
independently of the Hamiltonian parameters, which change the ground-state energy and the level spacing.

\begin{figure}[!h]
\begin{center}
\includegraphics[width=0.48\textwidth,angle=0]{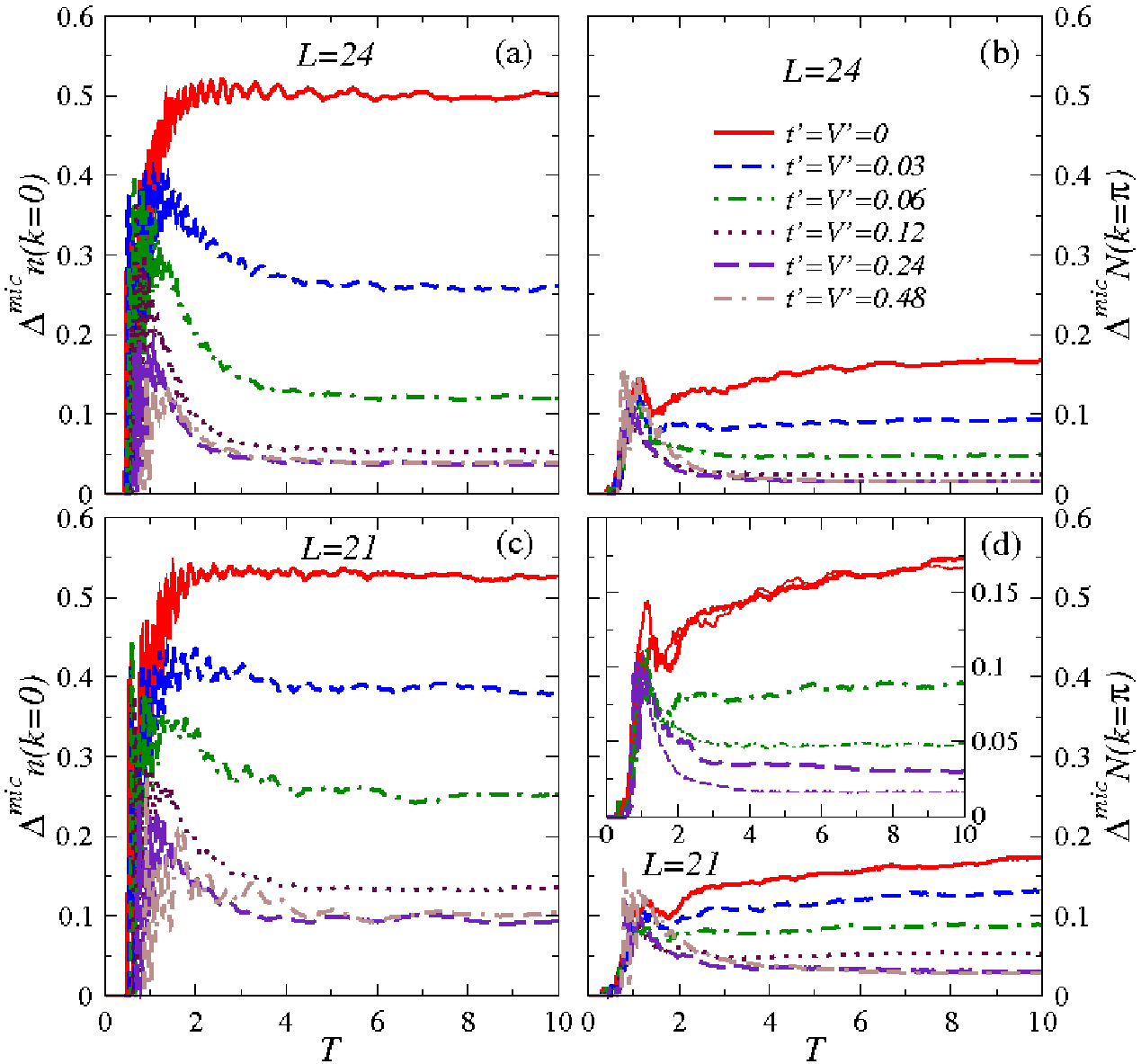}
\end{center}
\vspace{-0.6cm}
\caption{\label{Fig:DeviationfromETH} (Color online)
Average relative deviation of eigenstate expectation values with 
respect to the microcanonical prediction vs $T$ (see text). 
Results are presented for: (a),(c) $\Delta^{mic}n(k=0)$ and 
(b),(d) $\Delta^{mic}N(k=\pi)$ in systems with: 
(a),(b) $L=24$, $N_f=8$ and (c),(d) $L=21$, $N_f=7$. 
The inset in (d) shows a direct comparison of $\Delta^{mic}N(k=\pi)$ 
between the systems with $L=24$, $N_f=8$ (thin lines) and
the systems with $L=21$, $N_f=7$ (thick lines). The values
of $t',V'$ in the three cases depicted in the inset are, from top to bottom,
$t'=V'=0$, $t'=V'=0.06$, and $t'=V'=0.24$, and follow the same 
legends of the main panel shown in (b). In all cases $t=V=1.0$.}
\end{figure}

In Fig.\ \ref{Fig:DeviationfromETH}, we present results for $\Delta^{mic}n(k=0)$ [(a),(c)] and 
$\Delta^{mic}N(k=\pi)$ [(b),(d)] vs $T$ for six different values of $t'=V'$ in systems
with 24 lattice sites and eight fermions [(a),(b)] and 21 lattice sites and seven fermions [(c),(d)].
The average relative deviations for both observables are presented in the same scale, which 
immediately allows one to see what was already evident in Fig.\ \ref{Fig:ETH}, namely, that the 
average relative deviations for $\Delta^{mic}n(k=0)$ are larger (more than two times larger) than 
the corresponding ones for $\Delta^{mic}N(k=\pi)$ for any given value of $t',V'$. For temperatures 
$T\gtrsim 1.5$, one can see that those deviations for both observables saturate with increasing 
$t',V'$ for $t'=V'>0.12$. Comparing the results for the same observables but for different systems 
sizes, one can see that with increasing system size the average relative deviations are decreasing. 
This can be better seen in the inset in Fig.\ \ref{Fig:DeviationfromETH}(d), where we have presented 
a direct comparison of $\Delta^{mic}N(k=\pi)$ for three different values of $t',V'$ in the systems 
with 21 (thick lines) and 24 (thin lines) sites. At integrability, one does not see much of a change 
with changing system size for $\Delta^{mic}N(k=\pi)$, but some reduction can be seen for 
$\Delta^{mic}n(k=0)$ [Figs.\ \ref{Fig:DeviationfromETH}(a) and \ref{Fig:DeviationfromETH}(c)]. For
$t'=V'>0$, we do see a clear reduction in all cases. At low temperatures $T<1.5$, the density 
of states is low and in many cases the use of the microcanonical ensemble is not justified. From 
Fig.\ \ref{Fig:ETH}, one can see that in that regime (of low energies) the fluctuations of 
the observables are very large and thermalization is not expected to occur. An important 
question that needs to be addressed in the future is what will happen with that window of 
temperatures (energies) with increasing the system size. For our small systems, we do not
see a noticeable change with increasing system size. 

\begin{figure}[!h]
\begin{center}
\includegraphics[width=0.36\textwidth,angle=0]{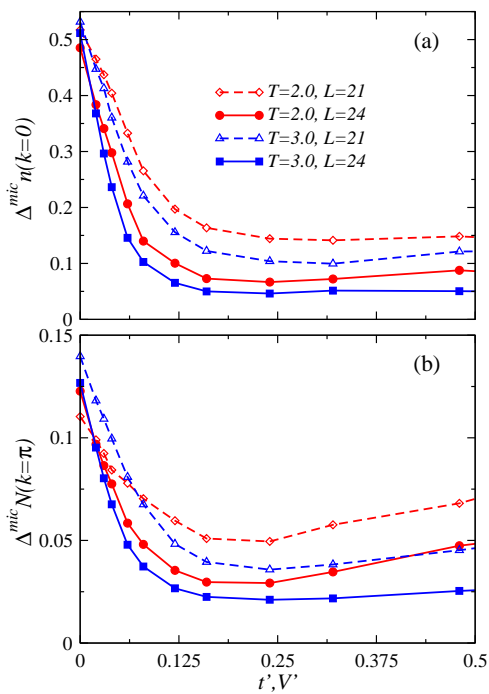}
\end{center}
\vspace{-0.6cm}
\caption{\label{Fig:DeviationfromETH_tpVp} (Color online)
Average relative deviation of eigenstate expectation values with respect to the 
microcanonical prediction at two fixed temperatures ($T=2.0$ and $T=3.0$) and for two system 
sizes ($L=21$, $N_f=7$, and $L=24$, $N_f=8$) vs $t',V'$. Results are presented for: 
(a) $\Delta^{mic}n(k=0)$ and (b) $\Delta^{mic}N(k=\pi)$. In all cases $t=V=1.0$.}
\end{figure}

An alternative way to present some of the results depicted in 
Fig.\ \ref{Fig:DeviationfromETH}, which can help us make direct contact with 
the results discussed in Fig.\ \ref{Fig:Thermodynamics}, is to take two values
of the effective temperature within the microcanonical ensemble and plot the 
average relative deviation of eigenstate expectation values with respect 
to the microcanonical prediction for those temperatures 
as a function of the integrability breaking parameters $t',V'$. 
This is done in Fig.\ \ref{Fig:DeviationfromETH_tpVp} for $T=2.0$ and $T=3.0$, and 
$L=21$ and $L=24$. Figure \ref{Fig:DeviationfromETH_tpVp} clearly shows that the breakdown 
of thermalization seen in Fig.\ \ref{Fig:Thermodynamics} as one approaches integrability
is directly related to the increase in the relative deviation of eigenstate 
expectation values with respect to the microcanonical prediction, i.e., to the increase 
in the fluctuations of the eigenstate to eigenstate expectation values of $\hat{n}(k)$ 
and $\hat{N}(k)$, or what is the same to the failure of ETH. A comparison between 
Figs.\ \ref{Fig:DeviationfromETH_tpVp}(a) and  \ref{Fig:DeviationfromETH_tpVp}(b) also
helps in understanding why in Fig.\ \ref{Fig:Thermodynamics} the difference between
the diagonal and microcanonical ensembles is larger for $n(k)$ than for $N(k)$, even
when one is far from integrability. Figure \ref{Fig:DeviationfromETH_tpVp}(a) shows
that for the effective temperatures selected in Fig.\ \ref{Fig:Thermodynamics}, 
$\Delta^{mic}n(k=0)$ is more than two times larger than $\Delta^{mic}N(k=\pi)$. 
The reduction in $\Delta^{mic}n(k=0)$ and $\Delta^{mic}N(k=\pi)$, with increasing
system size, can also be seen in Fig.\ \ref{Fig:DeviationfromETH_tpVp} when comparing 
the results for $L=21$ and $L=24$.

\section{Off-diagonal matrix elements}
\label{Sec:OffDiagonal}

While the diagonal elements of the observables, in the basis of the eigenstates of the 
Hamiltonian, helped us understand whether after relaxation the expectation values
of generic few-body observables can be described by standard statistical mechanical 
approaches, it follows from Eqs.\ (\ref{Eq:timeevolution}) and (\ref{Eq:diagonal}) that
the off-diagonal elements of the observables can help us understand the 
relaxation dynamics and temporal fluctuations of observables after relaxation,
\begin{equation}
\langle \widehat{O}(\tau) \rangle - \overline{\langle \widehat{O}(\tau) \rangle}
=\sum_{\substack{\alpha\beta \\ \alpha\neq\beta}} C_{\leftindex}^{\star} C_{\rightindex}^{}
e^{i(E_{\leftindex}-E_{\rightindex})\tau}
O_{\leftindex\rightindex}\, .
\label{Eq:timefluctuations}
\end{equation}

\begin{figure}[!h]
\begin{center}
\includegraphics[width=0.48\textwidth,angle=0]{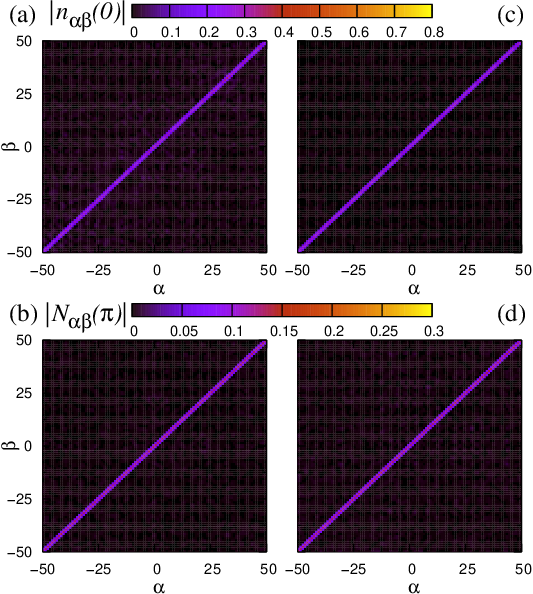}
\end{center}
\vspace{-0.6cm}
\caption{\label{Fig:OffDiagonaltpVp0.32_L24T3.0} (Color online)
Off-diagonal elements of $\hat{n}(k=0)$ [(a),(c)] and $\hat{N}(k=\pi)$ [(b),(d)] for
fermions [(a),(b)] and hardcore bosons [(c),(d)] far from integrability. Results are shown 
for the 100 eigenstates closest to the one with energy $E=-4.26$ ($T=3.0$, see text) 
for the fermions [(a),(b)] and $E=-4.62$ ($T=3.0$, see text) for the hardcore bosons [(c),(d)]. 
In all cases $t=V=1.0$, $t'=V'=0.32$, $L=24$, and $N_f=N_b=8$.}
\end{figure}

As we pointed out in Sec.\ \ref{Sec:dynamicsF}, one of our findings in this work is that
the momentum distribution function of fermions exhibits a slower relaxation dynamics than
the one seen for the same observable in an identical system of hardcore bosons 
\cite{rigol09STATa}. These differences were seen far from the integrable point as well 
as at the integrable point. In Figs.\ \ref{Fig:OffDiagonaltpVp0.32_L24T3.0}(a) and 
\ref{Fig:OffDiagonaltpVp0.32_L24T3.0}(c), we compare the off-diagonal elements of 
$\hat{n}(k=0)$ between a system of spinless fermions [Fig.\ \ref{Fig:OffDiagonaltpVp0.32_L24T3.0}(a)]
and an identical system with hardcore bosons [Fig.\ \ref{Fig:OffDiagonaltpVp0.32_L24T3.0}(c)], 
both systems are far from integrability [with $t'=V'=0.32$] and the central eigenstate has
an energy for which the temperature of a canonical ensemble with the same energy 
would be $T=3.0$. It is apparent in these two figures that the off-diagonal 
elements of $\hat{n}(k=0)$ are larger for the fermions than for the hardcore bosons, 
in particular close to the diagonal.

The relaxation dynamics of our other observable of interest $N(k)$, on the other hand, 
was seen to be very similar between the fermions studied in this work and the hardcore 
bosons studied in Ref.\ \cite{rigol09STATa}. In both systems, this observable relaxed very
quickly (in a time scale $\tau\sim 1/t$). Figures \ref{Fig:OffDiagonaltpVp0.32_L24T3.0}(b) 
and \ref{Fig:OffDiagonaltpVp0.32_L24T3.0}(d) depict results for the off-diagonal
elements of $\hat{N}(k=\pi)$. The results for fermions and hardcore bosons are almost 
indistinguishable in this case, with all off-diagonal elements being much smaller than 
the diagonal ones. The contrast between the magnitude of the off-diagonal elements of 
$\hat{n}(k=0)$ in Fig.\ \ref{Fig:OffDiagonaltpVp0.32_L24T3.0}(a) and $\hat{N}(k=\pi)$ 
in Fig.\ \ref{Fig:OffDiagonaltpVp0.32_L24T3.0}(b) also explains the difference between 
the relaxation dynamics of both observables in fermionic systems depicted in 
Fig.\ \ref{Fig:TimeEvolution_L24T2.0}.

\begin{figure}[!h]
\begin{center}
\includegraphics[width=0.48\textwidth,angle=0]{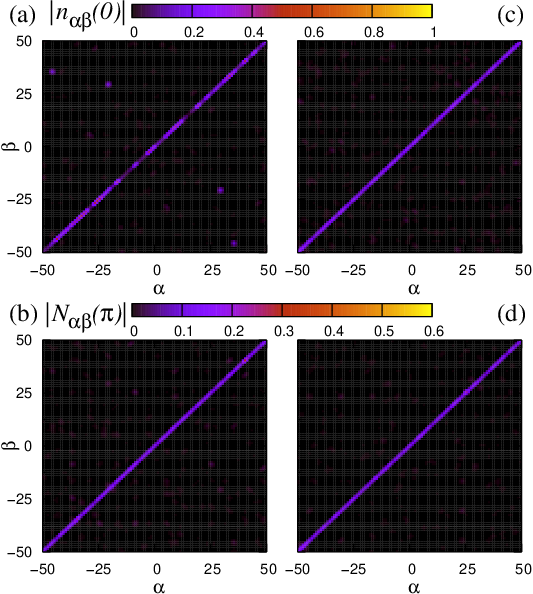}
\end{center}
\vspace{-0.6cm}
\caption{\label{Fig:OffDiagonaltpVp0.00_L24T3.0} (Color online)
Off-diagonal elements of $\hat{n}(k=0)$ [(a),(c)] and $\hat{N}(k=\pi)$ [(b),(d)] for
fermions [(a),(b)] and hardcore bosons [(c),(d)] at integrability. Results are shown for 
the 100 eigenstates closest to the one with energy $E=-3.84$ ($T=3.0$, see text) 
for the fermions [(a),(b)] and $E=-3.84$ ($T=3.0$, see text) for the hardcore bosons [(c),(d)]. 
In all cases $t=V=1.0$, $t'=V'=0.00$, $L=24$, and $N_f=N_b=8$.}
\end{figure}

Results for the same quantities for fermions and hardcore bosons at the integrable point
are shown in Fig.\ \ref{Fig:OffDiagonaltpVp0.00_L24T3.0}. We find the results at integrability
in stark contrast with those far away from the integrable point. At integrability, 
the off-diagonal elements of both observables are in most cases very small or zero, but then 
there are some states between which the off-diagonal elements can be very large (much larger than any 
off-diagonal element seen in the nonintegrable case). In addition, very large off-diagonal elements 
can be seen between states that have quite different energies. In 
Fig.\ \ref{Fig:OffDiagonaltpVp0.32_L24T3.0}, 
the largest off-diagonal elements in the nonintegrable case always occur between states that 
are close in energies, and they are seen to reduce as the energy of the states depart from 
each other. This is not the case at integrability.

\begin{figure}[!h]
\begin{center}
\includegraphics[width=0.48\textwidth,angle=0]{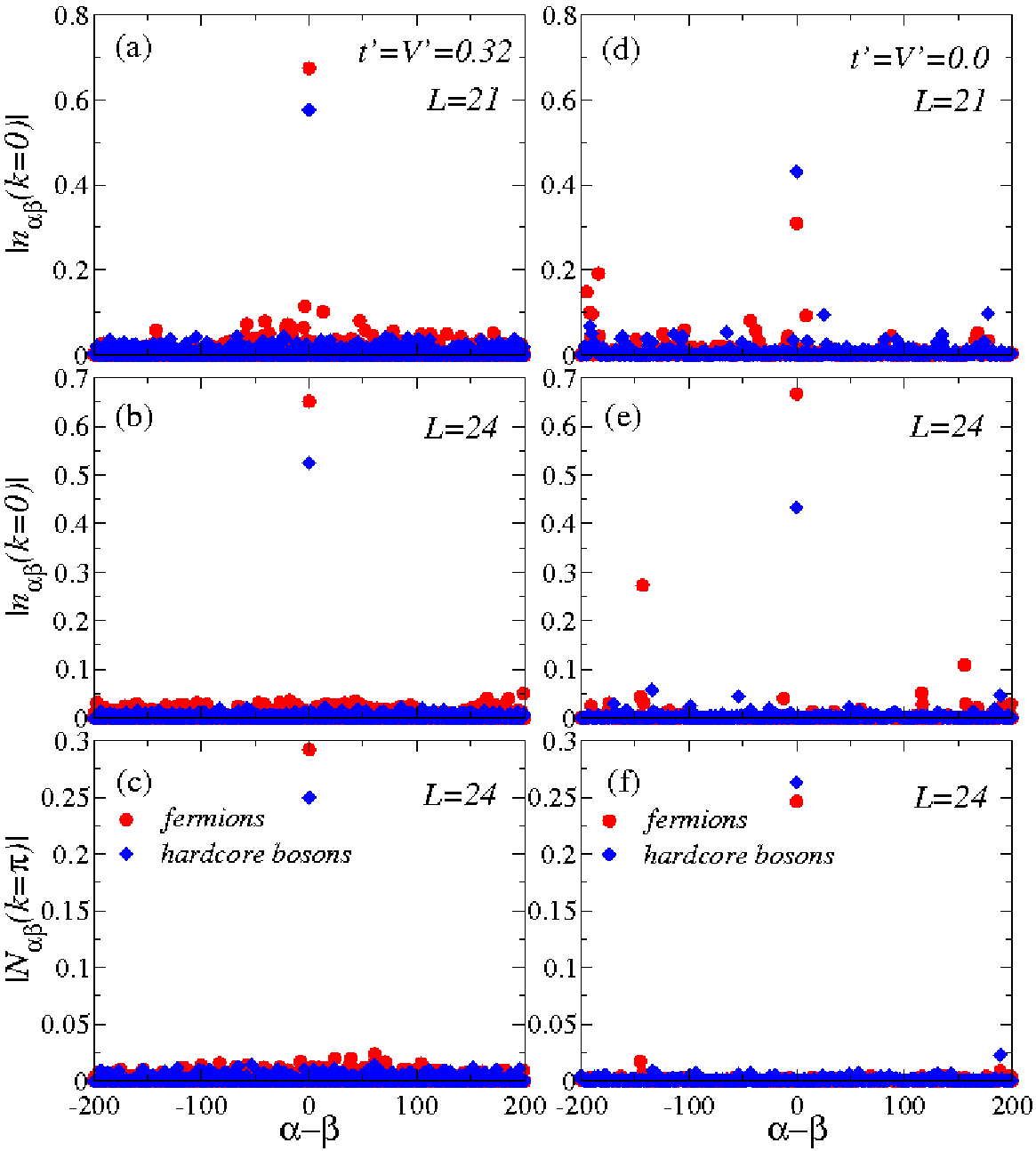}
\end{center}
\vspace{-0.6cm}
\caption{\label{Fig:OffDiagonal_ALL} (Color online)
Off-diagonal elements of $\hat{n}(k=0)$ [(a),(b),(d),(e)] and $\hat{N}(k=\pi)$ [(c),(f)] 
for fermions and hardcore bosons away from integrability [(a)--(c)] and at integrability [(d)--(f)]. 
Panels (b),(c),(e),(f) correspond to cuts across Figs.~\ref{Fig:OffDiagonaltpVp0.32_L24T3.0} and 
\ref{Fig:OffDiagonaltpVp0.00_L24T3.0}, with either $\alpha=0$ or $\beta=0$. Panels (a) and (d) depict 
results for the off-diagonal elements of $\hat{n}(k=0)$ in systems with $L=21$. In all cases, the eigenstate 
defining the center of the window has the energy closest to that of a canonical ensemble with temperature $T=3.0$.}
\end{figure}

A more quantitative understanding of these issues can be gained by taking one state from
Figs.\ \ref{Fig:OffDiagonaltpVp0.32_L24T3.0} and \ref{Fig:OffDiagonaltpVp0.00_L24T3.0} and
plotting the off-diagonal elements between that state (which we take to be the one at the 
center) and all the closest energy eigenstates for our two observables of interest. These are
depicted in Fig.\ \ref{Fig:OffDiagonal_ALL} for the nonintegrable case with 
$t'=V'=0.32$ [(a)--(c)] and at integrability $t'=V'=0$ [(d)--(f)].

First, let us focus on the behavior of $n_{\alpha\beta}(k=0)$. Results for that quantity
in the nonintegrable fermionic and hardcore boson cases are shown in panels (a) and (b) 
for two lattices with $L=21$ and $L=24$, respectively. For each lattice size, one can 
clearly see that the off-diagonal elements of $\hat{n}(k=0)$ of the fermions are always
larger than the ones of the bosons, and the differences are largest close to the diagonal. 
As the energies between the states depart from each other $n_{\alpha\beta}(k=0)$ for bosons 
and fermions become similar. The comparison between the two lattices also shows that 
with increasing system size the off-diagonal elements of the observables decrease, but the 
fact that $n_{\alpha\beta}(k=0)$ for the fermions is larger close to the diagonal remains. 
Similarly, we find that with decreasing the energy of the state the off-diagonal elements 
of $n_{\alpha\beta}(k=0)$ also become larger, although for temperatures $T>1.0$ this effect
is smaller than the finite-size effect difference between panels (a) and (b) 
in Fig.\ \ref{Fig:OffDiagonal_ALL}.

Results for $n_{\alpha\beta}(k=0)$ at integrability are presented in panels (d) and (e) of
Fig.\ \ref{Fig:OffDiagonal_ALL}. There one can also see that the values of $n_{\alpha\beta}(k=0)$
are in general larger for fermions than for hardcore bosons and that they decrease with increasing
the system size. However, here fermions and bosons share a common feature, namely, large 
off-diagonal elements can be seen between states that have energies that are increasingly
different from each other. This, in conjunction with the fewer number of states that have a 
significant overlap with the initial state, can lead to larger temporal fluctuations during 
the time evolution after a quench at integrability. This effect is enhanced at lower temperatures 
where we see some increase in the magnitude of the off-diagonal elements and a decrease in the 
number of states that overlap with the initial state (see Fig.\ \ref{Fig:StateCount}).

Finally, the off-diagonal elements for $\hat{N}(k=\pi)$ far from integrability and at 
integrability are presented in Fig.\ \ref{Fig:OffDiagonal_ALL}, panels (c) and (f), 
respectively. Those panels show that, for the density-density structure factor, off-diagonal 
elements are always much smaller than the diagonal ones, and they are relatively smaller 
than the ones of $\hat{n}(k)$. At integrability 
[Fig.\ \ref{Fig:OffDiagonal_ALL}(f)], one can see a few values of $N_{\alpha\beta}(k=\pi)$ 
that are clearly larger than the rest, but they are still less and relatively smaller than 
the ones seen for $n_{\alpha\beta}(k=0)$ in Fig.\ \ref{Fig:OffDiagonal_ALL}(e). These results
explain why $\hat{N}(k)$ takes less time to relax to the diagonal ensemble prediction 
and why the temporal fluctuations after relaxation are also smaller for this observable. 

\section{Summary}
\label{Sec:summary}

We have presented a detailed study of the relaxation dynamics after a quantum quench and the 
description after relaxation of the momentum distribution function $n(k)$ (a one-particle 
observable) and the density-density structure factor $N(k)$ (a two-particle observable) 
of spinless fermions with nearest and next-nearest hoppings and interactions in 
one-dimensional periodic lattices. We have also compared some results for fermions with 
those of a similar hardcore boson system. We should stress that those models are typical 
for describing the physics of one-dimensional systems. For example, they can be mapped onto
the spin-1/2 $XXZ$ linear chain with next-nearest-neighbor interactions. (The observables
analyzed here are related to the different spin-spin correlation functions.) Hence, we expect
the results reported in this manuscript to be generic and apply to other gapless models and 
observables in one-dimensional systems.

We have shown that, in general, $n(k)$ for fermions exhibits a slower relaxation dynamics, 
i.e., it takes longer to relax to an equilibrium distribution than other observables for fermions 
[such as $N(k)$] and than $n(k)$ for a similar one-dimensional system of hardcore bosons (studied in 
Ref.\ \cite{rigol09STATa}). Close to and at integrability, we have also found that $n(k)$ 
may even fail to relax to an equilibrium distribution while $N(k)$, and $n(k)$ for a similar
bosonic system, do relax to equilibrium. We have shown that this behavior of the dynamics of 
$n(k)$ is related to the particularly large off-diagonal elements of $\hat{n}(k)$ in the 
basis of the eigenstates of the Hamiltonian. Those off-diagonal elements are larger than the 
ones for the same observable in hardcore boson systems, and much larger than the ones for 
$\hat{N}(k)$ in the same fermionic system. From the contrast between the dynamics of $n(k)$ 
and $N(k)$ emerges a general picture in which some few-body observables in isolated quantum 
systems may quickly relax to an equilibrium distribution while other observables may take 
longer to relax, or even fail to relax, to an equilibrium distribution.

We have shown that far from integrability observables after relaxation are well described 
by standard ensembles from statistical mechanics, such as the microcanonical ensemble, which
is particularly relevant to our small and isolated systems. The ability of the microcanonical 
ensemble to predict the expectation value of few-body observables after relaxation was shown
to be related to the validity of the eigenstate thermalization hypothesis, within which the 
expectation values of observables in the eigenstates of the Hamiltonian already exhibit 
thermal behavior. We have also shown that as one approaches the integrable point 
thermalization breaks down, with the differences between the microcanonical predictions and 
the results of the relaxation dynamics increasing continuously as one converges toward the
integrable point. We have established that there is a clear correlation between the failure
of the system to thermalize and the failure of the eigenstate thermalization hypothesis, i.e.,
as one approaches integrability the differences between the eigenstate expectation values of
few-body observables increases between eigenstates that are close in energy, and this happens
over the full spectrum of the Hamiltonian.

Our results here have been obtained for small one-dimensional lattices, which are relevant
to current experiments in one-dimensional geometries \cite{paredes04}. However, several 
important questions remain to be addressed in future works. For example, what happens to the point 
at which thermalization breaks down (at all temperatures) as one increases the system size. 
Far from integrability, we find that for our small systems thermalization occurs whenever the 
effective temperature is $T\gtrsim 1.5$ \cite{temperature}. 
Another question that needs to be addressed is what happens with the window of energies where 
thermalization fails to occur far from integrability as one increases the system size. 
Finally, there is the question of the time scale that $n(k)$ of the fermions takes to relax 
to the thermal distribution, and the emergence of intermediate quasistationary distributions 
\cite{moeckel08,moeckel09}, as the size of the system is increased. Experiments with 
ultracold gases will generate many new questions and help address the current ones.

\begin{acknowledgments}
This work was supported by the U.S. Office of Naval Research under Award No.\ N000140910966
and by a Summer Academic Grant from Georgetown University. We thank Amy Cassidy and 
Itay Hen for useful comments on the manuscript. We are grateful to the Aspen 
Center for Physics for hospitality.
\end{acknowledgments}


\begin{thebibliography}{33}
\bibitem{temperature} By the effective temperature ($T$), we mean the temperature of a 
canonical ensemble with the same energy as the (isolated) time-evolving 
state. This temperature help us gauge the excitation energy of the time-evolving 
state with respect to the ground state. In addition, in thermodynamics systems 
where observables are expected to thermalize, $T$ allows one to predict the expectation 
values of observables of interest after relaxation.
\end{thebibliography}
\end{document}